\documentclass[aps,prl,twocolumn,superscriptaddress, showpacs]{revtex4}
\usepackage{color}
\usepackage{amsmath}
\usepackage{amsfonts}
\usepackage{amssymb}
\usepackage{graphicx}

\begin{document}

\title{Hole boring in a DT pellet and fast ion ignition with ultra-intense laser pulses}
\author{N. Naumova}
\affiliation{Laboratoire d'Optique Appliqu\'{e}e, ENSTA, Ecole Polytechnique, CNRS, 91761 Palaiseau, France}
\author{T. Schlegel}
\affiliation{Helmholtzzentrum f\"ur Schwerionenforschung GmbH, Planckstrasse 1, D-64291 Darmstadt, Germany}
\author{V. T. Tikhonchuk}
\affiliation{Centre Lasers Intenses et Applications, Universit\'e Bordeaux 1 - CEA - CNRS, 33405 Talence Cedex, France}
\author{C. Labaune}
\affiliation{Laboratoire pour l'Utilisation des Lasers Intenses, CNRS - CEA - Ecole Polytechnique - Universit\'e Pierre et Marie Curie, 91128 Palaiseau Cedex, France}
\author{I. V. Sokolov}
\affiliation{Space Physics Research Laboratory, University of Michigan, Ann Arbor, MI 48109, USA}
\author{G. Mourou}
\affiliation{Laboratoire d'Optique Appliqu\'{e}e, ENSTA, Ecole Polytechnique, CNRS, 91761 Palaiseau, France}
\date{\today}

\begin{abstract}
Recently achieved high intensities of short laser pulses open new prospects in their application to hole boring in inhomogeneous overdense plasmas and for ignition in precompressed DT fusion targets. A simple analytical model and numerical simulations demonstrate that pulses with intensities exceeding $10^{22}$\,W/cm$^2$ may penetrate deeply into the plasma as a result of efficient ponderomotive acceleration of ions in the forward direction. The penetration depth as big as hundreds of microns depends on the laser fluence, which has to exceed a few tens of GJ/cm$^2$. The fast ions, accelerated at the bottom of the channel with an efficiency of more than 20\%, show a high directionality and may heat the precompressed target core to fusion conditions.
\pacs{PACS: 52.50.Jm, 52.75.Di, 25.20.-x}
\end{abstract}
\maketitle

A progress in short-pulse lasers nowadays allows one to achieve focal
intensities larger than $10^{20}$W/cm$^2$, where the radiation
pressure becomes the dominant effect in driving the particle motion
\cite{wilks, denavit, esirkepov}. The plasma electrons are pushed
steadily by this force, and ions are accelerated in the strong
electrostatic field forming a shocklike structure
\cite{macchi,sorasio}. The use of circularly polarized laser light
might further improve the efficiency of the ponderomotive ion
acceleration avoiding strong electron overheating. It allows one to obtain a quasi-monoenergetic ion bunch in a homogeneous medium by adjusting the laser pulse and plasma parameters~\cite{esirkepov, macchi, sorasio, zhang1, klimo, sergeev, robinson}.

In this Letter, we investigate the effect of radiative ion
acceleration for the first time in inhomogeneous overcritical plasmas
with densities rising up to more than $100\,n_c$. The critical
electron density $n_c$ for laser wavelength $\lambda=0.8\,\mu$m
equals $1.72 \times 10^{21}$\,cm$^{-3}$, which corresponds to the mass
density for a deuterium-tritium (DT) plasma of 7.2\,mg/cm$^3$. We show
the possibility of hole boring in such plasmas driven by laser pulses
with intensities exceeding  $10^{22}$\,W/cm$^2$, which is much deeper
and more stable than considered before \cite{pukhov}. Following the
idea of fast ignition in thermonuclear fusion~\cite{tabak}, we propose
to use fast ions from the outer target layers directly irradiated by
the laser, to ignite the precompressed fuel core. The plasma density
profile of the DT target at the stagnation moment of the
precompression phase was obtained in numerical
simulations~\cite{ribeyre}. It corresponds to the baseline all-DT
target of the HiPER project~\cite{hiper} that is expected to release
$\sim 10$~MJ with the energy gain $\sim60$. The electron density
region $(1-100)\,n_c$, where the hole boring process is supposed to take place, can be roughly approximated by an exponential profile with a spatial scale of $\sim 20\,\mu$m.
\begin{figure} [b]
\includegraphics[width=7.5cm]{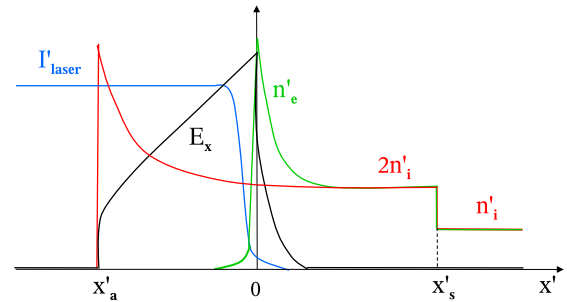}
\caption{Schematic structure of the piston and the electrostatic shock maintained by the radiation pressure. The frame is moving with the piston velocity. Curves: laser intensity, blue; electron density, green; electrostatic field, black; ion density, red.} \label{fig1}
\end{figure}

Simulation results described below demonstrate that the interaction of
an ultra-intense laser pulse with a high-density plasma takes the form
of a quasi-steady structure schematically shown in
Fig.\,\ref{fig1}. The laser ponderomotive potential sweeps all
electrons forward, so that the charge separation field forms a double
layer propagating with the velocity $v_f$, in which ions are
accelerated forward. The parameters of this double layer
(electrostatic shock or piston) standing in the range,
$[x_a^\prime,0]$, slowly evolve in time due to variations of the
plasma density and the laser intensity. The laser penetration is
bounded by a peak of the electron density. Much heavier ions penetrate
to the $x^\prime<0$ region, where they are reflected at $x_a^\prime$ and accelerated by the electrostatic field in the forward direction. In front of the piston, the plasma is practically neutral - ions and electrons reflected from the piston are propagating in an unperturbed plasma with the same velocity. The front shock at $x_s^\prime =v_ft^\prime$ separates the reflected flow and the unperturbed plasma.

\begin{figure} [b]
\includegraphics[width=8cm]{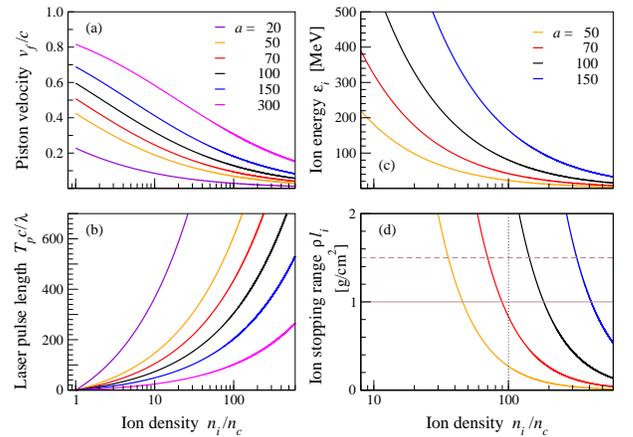}
\caption{(a) Piston velocity as a function of plasma density for a circularly polarized laser pulse. Each curve corresponds to the value of dimensionless vector potential $a= (I/ m_e n_c c^3)^{1/2}$, shown in the legend.
(b) Density dependence of time $T_p$ required for the laser pulse to
  traverse a plasma layer with an exponential density profile with $L
  = 20\,\mu$m, $n_{i\min}=n_c$, and $n_{i\max}=100 n_c$. The color of
  curves corresponds to the same field amplitudes as in (a). (c) Ion
  energy as a function of plasma density for several values of
  $a$. (d) Stopping range of ions with energies shown in (c) for the
  same density values. The dashed line represents the maximum areal density of the fusion pellet to be heated by ions. } \label{fig2}
\end{figure}

Let us consider the momentum conservation in the frame of reference co-moving with the piston. A Doppler-shifted laser beam is coming from the left and reflected back from the electron density spike. The momentum flux deposited by the photons is $2I^\prime/c$, where the light intensity in the piston frame $I^\prime$ is related to the incident laser intensity $I$ by the Lorentz transformation  $I^\prime = I\, (1-\beta_f)/(1+\beta_f)$. This light pressure has to be balanced by the pressure of the particles coming from the right side with the piston velocity  $v_f=\beta_f c$ (they are at rest in the laboratory frame) and will be elastically reflected back. The particle momentum flux reads, $2\,n_i^\prime v_f m_i \gamma_f v_f=2\,n_{i} m_i c^2\gamma_f^2 \beta_f^{\,2}$, where $\gamma_f=(1-\beta_f^2)^{-1/2}$ and $n_{i}$ is the upstream ion density in the laboratory frame. Here, we neglect the momentum of electrons because of their small mass. Then, by equating the photon and ion momentum fluxes, one finds the equation
\begin{equation}\label{e1}
  \frac{I}{n_{i} m_ic^3 }\,\frac{1-\beta_f}{1+\beta_f}=  \gamma_f^2 \beta_f^{\,2}
\end{equation}
for the piston velocity~\cite{zhang}. It can be solved explicitly:
\begin{equation}\label{e2}
  \beta_f = B/(1+ B),
\end{equation}
where $B= (I/n_{i}m_ic^3)^{1/2}$. In particular, for $B \ll 1$, the
motion of the piston is non-relativistic, and one finds the known
formula \cite{wilks, macchi}: $\beta_f \simeq B$. The piston velocity
characterizes also the reflection coefficient of the laser light $R=
(1-\beta_f)/(1+\beta_f)$. The piston slows down while propagating into
denser plasma regions also resulting in a decrease in laser absorption: $A=1-R$. The dependence of the piston velocity on the plasma density~\eqref{e2} is shown in Fig.\,\ref{fig2}\,(a).

The velocity of ions reflected from the piston in the laboratory frame is related to $v_f$ by the Lorentz transformation $v_i=2\beta_f c/(1+\beta_f^{2})$. Accordingly, the energy of accelerated ion  reads:
\begin{equation}\label{e3}
   \varepsilon_i = 2 m_ic^2 \gamma_f^2 \beta_f^{\,2}.
\end{equation}
In contrast to previous works, let us now consider an inhomogeneous plasma with a slowly varying density profile.
Integrating the equation for the piston coordinate $dx_p/dt = v_f$
over the densities ranging from  $n_{i\min}$ to $n_{i\max}$, one finds
the time needed to push the plasma up to the density $n_{i\max}$,
$T_{b}= c^{-1} \int dn_i\,L/\beta_f n_i$, where $L(n_i)=n_i/(dn_i/dx)$
is the density scale length. In case of an exponential profile with a
constant scale length $L$,
\begin{equation}\label{e4}
  T_{b} = 2L\,\left(m_ic /I \right)^{1/2}\,\left(n_{i\max}^{1/2}-n_{i\min}^{1/2}\right) + L_p/c,
\end{equation}
where $L_p$ is the plasma layer thickness. The first term in Eq.~\eqref{e4} determines the laser pulse duration needed to penetrate the chosen plasma layer $T_p= T_{b} - L_p/c$. Its dependence on the ion density in the interval $(1 - 100)\,n_c$ is plotted in Fig.\,\ref{fig2}\,(b). The time of 100 laser periods corresponds to 270\,fs for the laser wavelength $0.8\,\mu$m, which is a relatively short time compared to the typical lifetime of the plasma channel. The latter is about 30\,ps long. It can be estimated dividing the channel diameter $\sim 10\,\mu$m by the characteristic sound velocity in the ambient plasma, $\lesssim 300\,\mu$m/ns, for a temperature of 1\,keV.

Now we can find the energy spectrum and the fluence of accelerated ions. Within the time interval $dt$, the piston passes through the distance $v_f dt$ and accelerates $dN_i=n_{0i}v_f dt$ ions. These ions are distributed over the energy interval $d\varepsilon = (d\varepsilon_i/dx)\,v_f dt$. Thus, we find the following formula for the ion energy spectrum:
\begin{equation}\label{e5}
  \frac{dN_i}{d\varepsilon}= \frac{IL}{2m_i^2c^5\beta_f^4\gamma_f^6(1+\beta_f)},
\end{equation}
where $\beta_f$ is related to the ion energy according to Eq.~\eqref{e3}. Integrating the ion energy over the density range $[n_{i\min},n_{i\max}]$ with a constant scale length $L$, we find an expression for the ion fluence from this plasma region: $F_i = \int \varepsilon_i\,n_i\,dx$. For sufficiently large densities, such that $\beta_f \ll 1$, we  find:
\begin{equation}\label{e6}
  F_i \simeq \frac{2IL}{c} \ln\frac{n_{i\max}}{n_{i\min}}.
\end{equation}

To verify our analytical model and to confirm the theoretical
possibility of hole boring in an inhomogeneous plasma using
superintense laser pulses we performed one-dimensional (1D) and 2D
simulations with a relativistic electromagnetic particle-in-cell (PIC)
code \cite{naumova}. The code does not include binary collisions
between particles, but it accounts for the electron radiation losses,
which are important at high laser intensities \cite{sokolov}. We
consider the interaction of a circularly polarized laser pulse with
the intensity $I = 4\times 10^{22}$\,W/cm$^2$ and wavelength
$\lambda=0.8\,\mu$m ($a=100$) with a deuterium plasma layer. The laser
pulse has a step-like envelope with a 2$\lambda$ rising edge. At the
instant $t=0$, the pulse passes the plasma boundary at $x=0$. The
plasma layer has an exponential profile raising from 5 up to
100$\,n_c$ over a distance of $L_p=60\,\lambda$ with the density scale
length $L=20\,\lambda$. This exponential layer is extended by a uniform plasma layer of a density of 100$\,n_c$ to avoid the reflection of accelerated particles.

\begin{figure} [h]
\includegraphics[width=7.5cm]{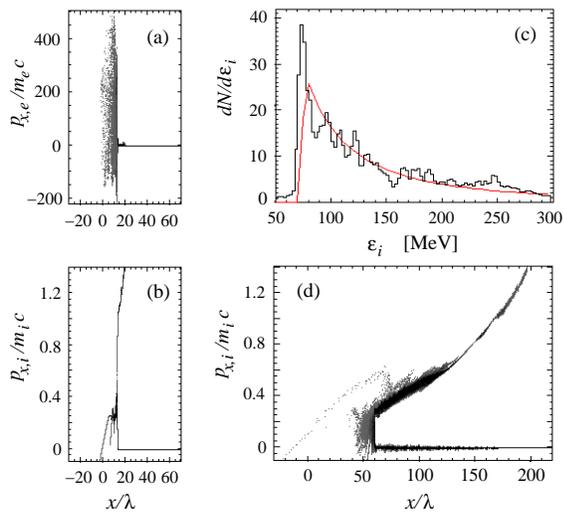}
\caption{Interaction of a circularly polarized laser pulse of
  intensity $4\times 10^{22}$ W/cm$^2$ with plasma having an
  exponential profile.  Momentum distribution $p_x(x)$ for (a)
  electrons and (b) ions at instant $t=30\,\lambda/c$. (c) Ion energy
  and (d) momentum distributions at the final instant
  $T_b=250\,\lambda/c$. The thin continuous line represents the analytical distribution function, Eq.~\eqref{e5}. Simulation parameters are given in the text. } \label{fig3}
\end{figure}

Figure~\ref{fig3}\,(a) shows the longitudinal momentum distribution $p_x(x)$ for electrons at instant $t=30\,\lambda/c$. The laser driven piston zone at $x=14\,\lambda$, where the laser pulse is reflected, can be easily identified. Most electrons are driven by the ponderomotive force of the laser field in the forward direction. However, due to the generation of a strong charge separation field, part of the cold electrons is accelerated backwards. The presence of electrons with positive and negative momenta in the piston zone demonstrates this fact. The backward accelerated electrons are interacting with the incident laser pulse, losing their energy by generating the radiation, and eventually reverse their motion \cite{sokolov}. For the given simulation parameters, $\sim1$\% of laser energy has been transformed to electrons and $\sim$10\% into high energy photons. The radiation losses prevent the electrons from escaping the piston zone and thus stabilize the laser propagation. The pulse duration $T_p=188\,\lambda/c$ needed to penetrate to the density 100\,$n_c$ estimated from Eq.~\eqref{e4} agrees well with the simulation.

Ions driven by the charge separation field  are accelerated and move forward as  shown in Fig.~\ref{fig3}\,(b). Because of relatively low electron energy, an almost complete neutrality is maintained in front of the piston. In a denser plasma, the piston speed decreases, which causes a decrease in ion energy observed in the momentum and energy distributions at the final time in Fig.~\ref{fig3}\,(c)-(d). The ion energy distribution agrees well with the analytical estimate following from Eq.~\eqref{e5} [thin line in Fig.~\ref{fig3}\,(c)]. The energy fluence carried with ions in the shown energy range $(50-300)$\,MeV amounts to 5.4\,GJ/cm$^2$, which is 27\% in respect to the total laser fluence $F_l= 20$\,GJ/cm$^2$.

\begin{figure} [h]
\includegraphics[width=7.5cm]{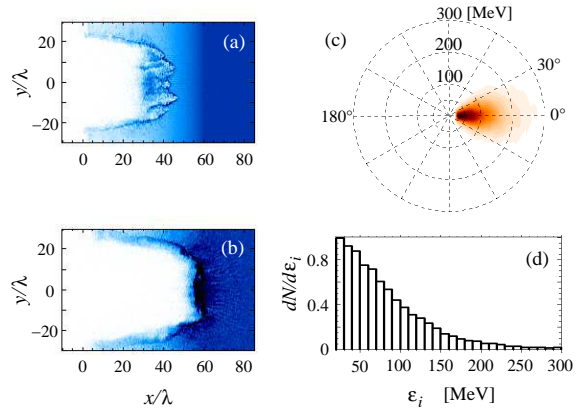}
\caption{Channel formation in plasma and ion acceleration in the laser
  piston regime. Ion density distribution at instants (a)
  $90\,\lambda/c$ and (b) $190\,\lambda/c$. (c) Angular energy
  distribution and (d) the energy distribution of accelerated ions in the domain $|y|\le 10\,\lambda$. Simulation parameters are the same as in Fig.~\ref{fig3}.} \label{fig4}
\end{figure}

Results of 2D PIC simulations are shown in Fig.~\ref{fig4}. They were performed for the same laser and plasma parameters as in the 1D case and with a flattop transverse laser intensity profile with a width of $20\lambda$ and exponential wings. The ion density distribution at two instants demonstrates an efficient hole boring in the plasma, a clean and a stable channel. The filamentation process is strongly suppressed due to radiation losses, similarly to the 1D case. The radiation process plays a positive role in the laser pulse channeling and ion acceleration, as it allows to maintain the electron thermal energy on a relatively low level and prevents the electron backward motion through the pulse \cite{sokolov}.

The final angular energy distribution and the energy distribution of
accelerated ions in the central region 
($|y|\le 10\,\lambda$) are plotted in Figs.~\ref{fig4}\,(c) and
\ref{fig4}(d), respectively. Compared to the 1D simulation, the hole boring velocity is slightly higher due to the pulse focusing in the channel. The major part of ions demonstrates a narrow angular distribution with the opening angle less than 6$^\circ$.

Let us determine the ion and laser parameters necessary to ignite the precompressed fuel in the pellet core. First, we deduce the ion characteristics. The scaling laws~\cite{atzeni} define the ignition energy $E_{ig}$ and the hot spot radius $r_{ig}$ as functions of the fuel density: $E_{ig} \simeq 18\,(\rho_0/\rho_{\max})^{1.85}$\,kJ and $r_{ig} \simeq 20\,(\rho_0/\rho_{\max})^{0.97}\,\mu$m, where $\rho_0=300$\,g/cm$^3$ is the reference density. Then one can estimate the energy flux needed to ignite the fuel $F_{ig}=E_{ig}/\pi r_{ig}^2$, which weakly depends on the fuel density:
\begin{equation}\label{e7}
  F_{ig} \simeq 1.4\,(\rho_{\max}/\rho_0)^{0.09}\,{\rm GJ/cm}^2.
\end{equation}
This energy must be carried by ions with the stopping range of the
order of the areal mass density in the compressed fuel $\rho_{\max}R
\simeq 1-1.5$ g/cm$^2$. Calculations of the ion stopping range with
the Monte-Carlo code TRIM \cite{trim} show that it can be interpolated
by a power law: $\rho\,l_i \simeq 10^{-3} \varepsilon_i^{1.8}$, where
$l_i$ is the ion stopping length. Here the ion energy is taken in MeV and the areal density $\rho\,l_i$ in g/cm$^2$. For the areal density of the compressed fuel $\sim 1.5$ g/cm$^2$, the maximum energy of fast ions fully stopped in the core is limited by the value $\sim 60$\,MeV.

The laser and plasma parameters that are needed to produce the ions with such an energy can be found from Fig.\,\ref{fig2}\,(c) and (d). The laser pulse with the amplitude $a = 70$ ($I \simeq 2 \times 10^{22}$\,W/cm$^2$) would accelerate ions to energies less than 60\,MeV in the plasma region with densities larger than $n_{i\min}=70\,n_c$. Laser pulses with smaller amplitudes $a<50$ are probably not appropriate for stable hole boring, because the long time $T_{b}$ implies a strong sensitivity on 2D effects. The laser pulse may lose its propagation stability, if the ponderomotive force is not strong enough to evacuate most of the electrons and to clean the channel in the underdense plasma~\cite{li}. Then, the Raman scattering would destroy the pulse and impede its access to the denser plasma layers.

In estimating the ion fluence, one has to account for the divergence
of the ion beam. Taking the density of compressed core $\rho_{\max}=400$ g/cm$^3$, we estimate the radius of ignition spot as $15\,\mu$m. Assuming the ion beam angular divergence of $\approx 6^\circ$ taken from Fig.\,\ref{fig4}\,(c) and a distance of propagation $\approx 50\,\mu$m \cite{ribeyre}, one finds that the beam radius increases by 5\,$\mu$m. Then one has to deliver the laser energy in the focal spot of $10\,\mu$m and to generate the ion beam with the fluence $F_i \simeq 2\,F_{ig}\simeq 3$\,GJ/cm$^2$. Applying Eq.\,\eqref{e6}, we obtain the upper density value $n_{i\max} \simeq 200\,n_c$ of the plasma acceleration layer with an exponential density profile $L = 20\,\mu$m. The ion energy spectrum generated in the density range from 70 to $200\,n_c$ extends from 15 to 60 MeV. The laser pulse duration required for the ion acceleration is 0.75 ps, according to Eq. \eqref{e4}. Consequently, for the intensity $2 \times 10^{22}$\,W/cm$^2$, the laser fluence is 15\,GJ/cm$^2$. Taking the focal spot area $3\times10^{-6}$~cm$^2$ we estimate the laser energy of 45\,kJ. This value is of the same order of magnitude as in other schemes of fast ignition, although the required power is 60\,PW. Such a high power is supposed to be achieved in the second stage of the ELI project~\cite{ELI}.

In addition, we have to account for the energy needed to assure laser propagation to the acceleration zone  $n_{i\min}=70\,n_c$ through the undercritical plasma. Here one can use a longer laser pulse with a lower intensity $\sim 10^{21}$\,W/cm$^2$. Applying the scaling law from \cite{li}, we estimate the pulse duration of 3\,ps necessary to reach the critical density. According to Eq.~\eqref{e4}, one needs 4\,ps more to reach the bottom of acceleration zone. Therefore, the hole boring requires the laser fluence of about 7\,GJ/cm$^2$, half of the acceleration amount. This value can be reduced by appropriate shaping of the laser pulse.

In conclusion, we have demonstrated analytically and with PIC simulations that there is a range of laser pulse intensities $\sim 10^{22}$~W/cm$^2$ that could be efficient for hole boring, \emph{in situ} ion acceleration, and fast ignition of precompressed thermonuclear targets. The advantage of this approach is that it does not need any additional arrangements and can be applied for spherical targets, which are well adapted to high repetition rate operation.

This work was coordinated by Institute Lasers and Plasmas; it is supported by the ANR under the contract BLAN07-3-186728 and by the Region Aquitaine, project 34293.


\begin{references}
\bibitem{zhang} The analogous relationship obtained in ~\cite{zhang1} coincides with Eq.~\eqref{e1}, if the piston motion is taken into account while evaluating the photon reflection.
\end{references}
\end{document}